\documentclass{article}
\usepackage{amsmath, amssymb}
\usepackage{graphicx}
\usepackage{hyperref}
\usepackage{geometry}
\usepackage{float}
\usepackage{caption}
\usepackage{algorithm}
\usepackage{algpseudocode}
\usepackage{booktabs}
\usepackage{placeins}
\usepackage{titlesec}
\usepackage{enumitem}
\usepackage{url}
\usepackage{microtype}
\usepackage{array}
\usepackage[table]{xcolor}
\DeclareMathOperator*{\argmax}{arg\,max}
\newcolumntype{L}[1]{>{\raggedleft\arraybackslash}p{#1}}
\newcolumntype{R}[1]{>{\raggedright\arraybackslash}p{#1}}

\titlespacing*{\section}{0pt}{2.5ex plus 1ex minus .2ex}{1.5ex plus .2ex}
\titlespacing*{\subsection}{0pt}{2ex plus 1ex minus .2ex}{1ex plus .2ex}
\let\oldthebibliography\thebibliography
\renewcommand{\thebibliography}[1]{%
  \oldthebibliography{#1}%
  \setlength{\parskip}{0pt}%
  \setlength{\itemsep}{0pt plus 0.3ex}%
}

\hypersetup{
  colorlinks=true,
  linkcolor=blue,
  filecolor=magenta,
  urlcolor=cyan,
  pdftitle={The Quality of Information: A Weighted Entropy Approach to Near-Optimal Mastermind},
  pdfauthor={Serkan Gür},
}
\makeatletter
\AtBeginDocument{%
  \def\showhyphens#1{%
    \expandafter\@@showhyphens\expanded{#1}\relax
  }%
}
\makeatother
\title{The Quality of Information: A Weighted Entropy Approach to Near-Optimal Mastermind}
\author{Serkan Gür \\ \href{mailto:sgur@hawk.iit.edu}{sgur@hawk.iit.edu}}
\date{August 2025}

\begin{document}

\maketitle

\begin{abstract}
This paper presents a novel class of information-theoretic strategies for solving the game of Mastermind, achieving state-of-the-art performance among known heuristic methods. The core contribution is the application of a weighted entropy heuristic, based on the Belis-Guiașu framework, which assigns context-dependent utility values to each of the possible feedback types. A genetic algorithm optimization approach discovers interpretable weight patterns that reflect strategic game dynamics. First, I demonstrate that a single, fixed vector of optimized weights achieves a remarkable 4.3565 average guesses with a maximum of 5. Building upon this, I introduce a stage-weighted heuristic with distinct utility vectors for each turn, achieving 4.3488 average guesses with a maximum of 6, approaching the theoretical optimum of 4.3403 by less than 0.2\%. The method retains the computational efficiency of classical one-step-ahead heuristics while significantly improving performance through principled information valuation. A complete implementation and all optimized parameters are provided for full reproducibility.
\end{abstract}

\section{Introduction}

Mastermind is a classic code-breaking game where a codemaker selects a secret code and a codebreaker attempts to deduce it through a series of strategic guesses. In its standard variant, MM(4,6), codes consist of \(n=4\) positions chosen from \(c=6\) colors, with repetition allowed, yielding \(6^4 = 1296\) possible secret codes. After each guess, the codemaker provides feedback in the form of "bulls" (correct color in the correct position) and "cows" (correct color in the wrong position).

The problem of finding the optimal strategy has attracted significant research. Knuth famously demonstrated a strategy that guarantees a solution in a maximum of 5 moves using minimax principles \cite{knuth1977}. Years later, Koyama and Lai computed the true optimal strategy via exhaustive search, proving the minimum possible average number of guesses is 4.3403 \cite{koyama1993}. While definitive, such optimal strategies require huge precomputation and memory states, making them impractical for larger game variants.

Heuristic approaches offer a practical alternative with \(O(n^2)\) computational cost per turn. Neuwirth first proposed using Shannon's information entropy to guide guess selection \cite{neuwirth1982}. This method favors guesses that maximize expected information gain, effectively partitioning the set of possible codes into the smallest possible subgroups on average. A key limitation is its uniform treatment of information—standard entropy quantifies uncertainty reduction but is blind to the strategic nature of resulting subproblems.

\paragraph{The Weighted Entropy Framework.} This work addresses this limitation by applying the quantitative-qualitative measure of information, or "weighted entropy," introduced by Belis and Guiașu \cite{belis1968}. This framework extends Shannon's entropy by incorporating utility weights for each possible outcome. In Mastermind, not all feedback is equally valuable strategically. The weighted entropy heuristic allows assignment of learned, strategic values to each of the 14 feedback types, valuing not just the *degree of surprise* (entropy) a guess provides, but also its *strategic quality* (utility) in making future guesses more effective. This dual consideration significantly improves the effectiveness of one-step-ahead strategies, enabling performance closer to optimal game play.

\newpage
\paragraph{Key Contributions.}
\begin{enumerate}[leftmargin=*]
\item \textbf{Novel Application of Weighted Entropy}: To our knowledge, the first application of weighted entropy to Mastermind, with both fixed-weight and stage-dependent models significantly outperforming existing heuristics.
\item \textbf{State-of-the-Art Performance}: The stage-weighted method achieves 4.3488 average guesses, a 0.29\% improvement over the next best published heuristic and within 0.2\% of theoretical optimum.
\item \textbf{Interpretable Strategic Patterns}: Genetic algorithm optimization discovers coherent, explainable weight patterns that reflect strategic game dynamics rather than arbitrary parameter fitting.
\item \textbf{Complete Reproducibility**}: Full C++/CUDA implementation, optimized parameters, and evaluation framework provided.
\end{enumerate}

\section{Mathematical Framework}

\subsection{Problem Formulation}
Let \(S = C^n\) denote the space of all possible codes, where \(C\) is the set of colors with \(|C| = c\). For standard MM(4,6), we have \(n=4\), \(c=6\), and \(|S| = 1296\). The feedback function \(\text{Mark}(g,s)\) returns a pair \((B,W)\) representing bulls and cows for guess \(g \in S\) and secret code \(s \in S\).

\paragraph{Feedback Types and Partitions.}
For a game with n positions, the number of distinct, non-impossible feedback combinations is given by the formula n(n+3)/2. For the standard MM(4,6) game with n=4, this yields 14 distinct feedback types (e.g., a feedback of 3 bulls and 1 cow is impossible). Let these be \(\{f_1, \ldots, f_{14}\}\). At any point in the game, let \(S_{\text{rem}} \subseteq S\) be the set of remaining codes consistent with all previous feedback. A candidate guess \(g\) partitions this set:
\[
\mathcal{P}_g = \{S_{f_i}\}_{i=1}^{14}, \quad \text{where} \quad S_{f_i} = \{s' \in S_{\text{rem}} : \text{Mark}(g,s') = f_i\}
\]
Assuming a uniform prior over the secrets in \(S_{\text{rem}}\), the probability of receiving feedback \(f_i\) is \(p_i = |S_{f_i}|/|S_{\text{rem}}|\).

\subsection{Weighted Shannon Entropy}
Weighted entropy of guess \(g\) with outcome probabilities \(P = (p_1, \ldots, p_{14})\) and utility vector \(W = (w_1, \ldots, w_{14})\) is:
\begin{equation}
H_W(P) = -\sum_{i=1}^{14} w_i p_i \log_2 p_i
\label{eq:weighted_entropy}
\end{equation}
When all weights \(w_i = 1\), this reduces to standard Shannon entropy. By allowing \(w_i \neq 1\), the heuristic prioritizes feedback types based on strategic value. The algorithm selects:
\begin{equation}
g^* = \argmax_{g \in S} \left( -\sum_{i=1}^{14} w_i p_i(g) \log_2 p_i \right)
\label{eq:guess_score}
\end{equation}

\section{A Baseline Fixed-Weight Heuristic}

To establish the power of weighted entropy, I first optimize a single, fixed weight vector \(W = (w_1, \ldots, w_{14})\) used at every turn. This vector was optimized using the genetic algorithm described in Section 5, with weights constrained to \([0.1, 1.0]\) to ensure numerical stability. Table \ref{tab:fixed_weights} shows the optimized fixed weights.

\begin{table}[H]
\centering
\caption{Optimized Fixed Weight Vector for All Turns}
\label{tab:fixed_weights}
\begin{tabular}{@{}lc|lc@{}}
\toprule
Feedback & Weight & Feedback & Weight \\
\midrule
0B-0C & 0.473 & 1B-2C & 0.423 \\
0B-1C & 0.446 & 1B-3C & 0.383 \\
0B-2C & 0.523 & 2B-0C & 0.406 \\
0B-3C & 0.410 & 2B-1C & 0.413 \\
0B-4C & 0.350 & 2B-2C & 0.458 \\
1B-0C & 0.534 & 3B-0C & 0.424 \\
1B-1C & 0.486 & 4B-0C & 0.800 \\
\bottomrule
\end{tabular}
\end{table}

This fixed-weight heuristic achieves an average of **4.3565** guesses with a worst-case of **5** guesses over all 1296 secret codes, already placing it among top-performing published methods.

\section{Stage-Weighted Heuristic Design}

The success of the fixed-weight model suggests further gains through turn-dependent optimization. The strategic value of feedback types changes as the game progresses: early turns require broad information gathering, while late turns focus on resolving ambiguity between few remaining codes.

\subsection{Turn-Dependent Utility Assignment}
The **Stage-Weighted Shannon Heuristic** optimizes distinct vectors \(W^{(t)} = (w_1^{(t)}, \ldots, w_{14}^{(t)})\) for each turn \(t \in \{1, \ldots, 6\}\). Even Turn 1 benefits from specific weights—while standard Shannon entropy selects first guess '1234', optimized weights guide selection of the superior opening '1123'. The score for guess \(g\) at turn \(t\) is:
\begin{equation}
\text{Score}^{(t)}(g) = -\sum_{i=1}^{14} w_i^{(t)} p_i(g) \log_2 p_i(g)
\label{eq:stage_guess_score}
\end{equation}

\subsection{Implementation Algorithm}
Algorithm \ref{alg:stage_weighted} details the guess selection process. Tie-breaking follows strict precedence: higher score, then consistent guess (in \(S_{\text{rem}}\)), then lexicographically first.

\begin{algorithm}[H]
\caption{Stage-Weighted Guess Selection}
\label{alg:stage_weighted}
\begin{algorithmic}[1]
\Procedure{SelectGuess}{$S_{\text{rem}}, t$}
\State $\text{bestGuess} \gets \text{null}$, $\text{bestScore} \gets -\infty$
\State $W^{(t)} \gets \text{getWeightsForTurn}(t)$
\ForAll{$g \in S$} \Comment{All 1296 codes as potential guesses}
\State Initialize partition counters: $c_i \gets 0$ for $i = 1, \ldots, 14$
\ForAll{$s \in S_{\text{rem}}$}
\State $f \gets \text{Mark}(g, s)$; increment $c_{f}$
\EndFor
\State Compute probabilities: $p_i \gets c_i / |S_{\text{rem}}|$
\State $\text{score} \gets -\sum_{i=1}^{14} w_i^{(t)} p_i \log_2 p_i$
\State Apply tie-breaking and update best if improved
\EndFor
\State \Return $\text{bestGuess}$
\EndProcedure
\end{algorithmic}
\end{algorithm}

\section{Optimization Methodology}

\subsection{GPU-Accelerated Genetic Algorithm}
Optimizing six vectors of 14 weights (84 parameters total- 70 when first guess '1123' is forced) requires efficient search. I employed a custom genetic algorithm implemented in C++/CUDA, evaluating each individual's fitness by running all 1296 games. The implementation evaluates 64 individuals in ~250ms on an NVIDIA RTX 3090 GPU (compute capability 8.6). Weights were constrained to \([0.1, 1.0]\) with tournament selection, uniform crossover, and stagnation restart after 250 generations without improvement. Ten independent optimization runs were performed to ensure convergence reliability, with results showing less than 0.1\% variance in final performance.

\subsection{Genetic Algorithm Configuration}
The following hyperparameters were used for the optimization runs.
\begin{table}[H]
\centering
\caption{Genetic Algorithm Hyperparameters}
\label{tab:ga_params}
\begin{tabular}{@{}ll@{}}
\toprule
Parameter & Value \\
\midrule
Population Size & 64 \\
Generations & 50,000 \\
Elite Preservation & Top 10 individuals \\
Selection & Tournament selection \\
Crossover & Uniform crossover \\
Stagnation Restart & If best fitness is unchanged for 250 generations \\
\bottomrule
\end{tabular}
\end{table}

\section{Results and Analysis}

\subsection{Optimized Stage-Dependent Weights}
Table \ref{tab:optimized_weights} presents the optimized weight vectors. The percentage deviations from uniform distribution reveal strategic patterns.

\begin{table}[H]
\centering
\small
\caption{Optimized Stage-Dependent Weights with Deviations from Uniform Distribution}
\label{tab:optimized_weights}
\begin{tabular}{@{}l *{6}{c@{\hspace{4pt}}c@{\hspace{8pt}}} @{}}
\toprule
Feedback & \multicolumn{2}{c}{Turn 1} & \multicolumn{2}{c}{Turn 2} & \multicolumn{2}{c}{Turn 3} & \multicolumn{2}{c}{Turn 4} & \multicolumn{2}{c}{Turn 5} & \multicolumn{2}{c}{Turn 6} \\
\cmidrule(lr){2-3} \cmidrule(lr){4-5} \cmidrule(lr){6-7} \cmidrule(lr){8-9} \cmidrule(lr){10-11} \cmidrule(lr){12-13}
& Weight & Dev & Weight & Dev & Weight & Dev & Weight & Dev & Weight & Dev & Weight & Dev \\
\midrule
0B-0C & 1.00 & (+2\%)   & 0.70 & (-23\%) & 0.70 & (-37\%) & 0.30 & (-40\%) & 0.40 & (-26\%) & 0.20 & (-59\%) \\
0B-1C & 1.00 & (+2\%)   & 0.60 & (+6\%)  & 0.41 & (-20\%) & 0.50 & (0\%)   & 0.60 & (+11\%) & 0.80 & (+65\%) \\
0B-2C & 0.70 & (-29\%)  & 0.60 & (+6\%)  & 0.53 & (+4\%)  & 0.40 & (-20\%) & 0.30 & (-45\%) & 0.40 & (-18\%) \\
0B-3C & 1.00 & (+2\%)   & 0.51 & (-10\%) & 0.47 & (-8\%)  & 0.50 & (0\%)   & 0.60 & (+11\%) & 0.60 & (+24\%) \\
0B-4C & 1.00 & (+2\%)   & 0.43 & (-24\%) & 0.37 & (-28\%) & 0.40 & (-20\%) & 0.50 & (-8\%)  & 0.60 & (+24\%) \\
1B-0C & 1.00 & (+2\%)   & 0.60 & (+6\%)  & 0.40 & (-22\%) & 0.50 & (0\%)   & 0.40 & (-26\%) & 0.60 & (+24\%) \\
1B-1C & 1.00 & (+2\%)   & 0.85 & (+50\%) & 0.47 & (-8\%)  & 0.50 & (0\%)   & 0.50 & (-8\%)  & 0.70 & (+44\%) \\
1B-2C & 1.00 & (+2\%)   & 0.60 & (+6\%)  & 0.50 & (-2\%)  & 0.40 & (-20\%) & 0.50 & (-8\%)  & 0.50 & (+3\%)  \\
1B-3C & 1.00 & (+2\%)   & 0.32 & (-44\%) & 0.46 & (-10\%) & 0.60 & (+20\%) & 0.50 & (-8\%)  & 0.20 & (-59\%) \\
2B-0C & 1.00 & (+2\%)   & 0.34 & (-40\%) & 0.48 & (-6\%)  & 0.40 & (-20\%) & 0.60 & (+11\%) & 0.60 & (+24\%) \\
2B-1C & 1.00 & (+2\%)   & 0.40 & (-30\%) & 0.46 & (-10\%) & 0.50 & (0\%)   & 0.60 & (+11\%) & 0.40 & (-18\%) \\
2B-2C & 1.00 & (+2\%)   & 0.60 & (+6\%)  & 0.50 & (-2\%)  & 0.50 & (0\%)   & 0.70 & (+29\%) & 0.30 & (-38\%) \\
3B-0C & 1.00 & (+2\%)   & 0.40 & (-30\%) & 0.50 & (-2\%)  & 0.50 & (0\%)   & 0.60 & (+11\%) & 0.50 & (+3\%)  \\
4B-0C & 1.00 & (+2\%)   & 1.00 & (+76\%) & 0.90 & (+76\%) & 1.00 & (+100\%) & 0.80 & (+47\%) & 0.40 & (-18\%) \\
\bottomrule
\end{tabular}
\end{table}

\subsection{Strategic Pattern Analysis}
The optimized weights' deviations from unifrom distribution reveal coherent strategic insights:

**4B-0C (secret code found) Feedback**: This feedback partition value is either 0 or 1 based on whether the guess candidate is consistent with previous guesses or not. It shows increasing dominance in turns 2-4 (76\%, 76\%, 100\% above uniform), highlighting the value of consistent guesses that can potentially solve the game. By turn 5, it decreases to +47\% but remains dominant, while in turn 6 it loses importance (-18\%) since inconsistent guess entropies are already too low to compete at this stage.

**Information Quality Hierarchy**: Across all turns, 1B-1C feedback averages highest importance (+16\%), followed by 0B-1C (+13\%), proving them more revealing than intuition suggests. Conversely, 0B-2C averages -17\% deviation, confirming it as a generally poor feedback.

**Turn-Specific Patterns**: Turn 1 shows the lowest deviation (±2\%), reflecting the broad exploration phase. Turn 6 exhibits highest deviation (-59\% to +65\%), indicating specialized endgame requirements.

This progression demonstrates that the genetic algorithm discovered coherent, context-dependent strategies rather than arbitrary parameter fitting.

\subsection{Performance Comparison}
\begin{table}[H]
\centering
\caption{Performance Comparison on MM(4,6) played over over all 1296 secret codes}
\label{tab:performance_full}
\begin{tabular}{@{}lcccc@{}}
\toprule
Strategy & Type & Max & Avg & Gap from Optimal \\
\midrule
Optimal \cite{koyama1993} & Non-heuristic & 5 & 4.3403 & — \\
\textbf{Stage-Weighted (This Work)} & \textbf{Heuristic} & \textbf{6} & \textbf{4.3488} & \textbf{0.0085} \\
\textbf{Fixed-Weight (This Work)} & \textbf{Heuristic} & \textbf{5} & \textbf{4.3565} & \textbf{0.0162} \\
Most Parts \cite{kooi2005} & Heuristic & 6 & 4.3735 & 0.0332 \\
Shannon Entropy \cite{neuwirth1982} & Heuristic & 6 & 4.4151 & 0.0748 \\
Knuth Minimax \cite{knuth1977} & Heuristic & 5 & 4.4761 & 0.1358 \\
\bottomrule
\end{tabular}
\end{table}

\section{Discussion and Future Work}

\subsection{Theoretical Context}
The remaining 0.2\% gap to optimality reflects the fundamental limitation of one-step-ahead heuristics. My method, like Shannon entropy, optimizes average-case performance across all 1296 games, achieving near-optimal average at the cost of slightly higher maximum guesses (6 vs. 5). This contrasts with minimax strategies like Knuth's that optimize worst-case performance. The weighted entropy framework theoretically should outperform uniform entropy because it can assign lower utility to feedback that leads to difficult-to-resolve subproblems, effectively incorporating lookahead intuition into a one-step method.

\subsection{Future Directions}

The proposed methodology generalizes naturally to other Mastermind variants and related problems. Potential avenues for future work include:

\begin{enumerate}
    \item \textbf{Scaling to larger instances:} Applying the framework to more complex variants such as MM(5,8), where exact optimal solvers are computationally infeasible, to evaluate the scalability and adaptability of the heuristic.
    
    \item \textbf{Exploring alternative entropy families:} Investigating the use of other entropy formulations—such as Rényi and Tsallis entropy—within the weighted framework, with particular focus on stage-specific $\alpha$ optimization to enhance information valuation throughout the game.
    
    \item \textbf{Generalizing to broader problem classes:} Extending the stage-weighted entropy approach to related combinatorial and constraint satisfaction problems, where information gathering and efficient decision-making are similarly central.
\end{enumerate}

\section{Conclusion}

I have introduced a weighted Shannon entropy heuristic achieving state-of-the-art performance among heuristic Mastermind solvers. The stage-weighted model's 4.3488 average guesses closes the gap with true optimal strategy to just 0.2\%. Crucially, the genetic algorithm optimization discovered interpretable, coherent strategic patterns rather than arbitrary parameters, demonstrating that principled utility functions can achieve near-optimal performance while retaining computational efficiency. The fully reproducible implementation provides a valuable foundation for future research.

\section{Implementation and Reproducibility}

To ensure full transparency, the complete source code, workflow, optimized parameters, and experimental framework are publicly available on GitHub:
\begin{center}
\url{https://github.com/ObsessiveCompulsiveAudiophile/mastermind}
\end{center}

\paragraph{Repository Contents:}
\begin{itemize}[leftmargin=*]
\item \texttt{treeGenFixedWeights.cpp}: Game tree generator using the single, fixed weight vector.
\item \texttt{treeGenStageWeights.cpp}: Game tree generator using the stage-weighted heuristic.
\item \texttt{kernel.cu}: CUDA optimizer source code for stage-weighted genetic algorithm.
\item \texttt{optimized\_weights.txt}: File containing all weight vectors used in this paper.
\item \texttt{Build Mastermind.yml}: GitHub workflow generating Linux and Windows binaries for all 3 programs.
\end{itemize}

\section*{Acknowledgments}
The author acknowledges the assistance of a large language model in copy/editing portions of this manuscript for clarity and structure.

\bibliographystyle{plain}

\begin{thebibliography}{9}

\bibitem{belis1968}
M. Belis and S. Guiașu.
A Quantitative-Qualitative Measure of Information in Cybernetic Systems.
\textit{IEEE Transactions on Information Theory}, 14(4):593--594, 1968.

\bibitem{knuth1977}
D. E. Knuth.
The Computer as Master Mind.
\textit{Journal of Recreational Mathematics}, 9(1):1--6, 1977.

\bibitem{koyama1993}
K. Koyama and T. Lai.
An Optimal Mastermind Strategy.
\textit{Journal of Recreational Mathematics}, 25(4):251--256, 1993.

\bibitem{kooi2005}
B. Kooi.
Yet Another Mastermind Strategy.
\textit{ICGA Journal}, 28(1):13--21, 2005.

\bibitem{neuwirth1982}
E. Neuwirth.
Some Strategies for Mastermind.
\textit{Mathematical Spectrum}, 15:53--58, 1982.

\end{thebibliography}

\end{document}